\documentclass{iopart}
\usepackage{iopams}

\usepackage{graphicx}
\usepackage{bm}

\newcommand{\be}{\begin{equation}}
\newcommand{\ee}{\end{equation}}

\begin{document}

\title{Ultra-cold Polarized Fermi Gases}

\author{Fr\'ed\'eric  Chevy$^1$ and Christophe Mora$^2$}

\address{$^1$ Laboratoire Kastler Brossel, \'Ecole normale sup\'erieure, Paris}
\address{$^2$ Laboratoire Pierre Aigrain, \'Ecole normale
sup\'erieure, Paris }

\eads{chevy@lkb.ens.fr}
\eads{mora@lpa.ens.fr}

\begin{abstract}
Recent experiments with ultra-cold atoms have demonstrated the
possibility of realizing experimentally fermionic superfluids with
imbalanced spin populations.  We discuss how these developments
have shed a new light on a half-century old open problem in
condensed matter physics, and raised new interrogations of their
own.
\end{abstract}

\section{Introduction}

Since its discovery nearly a century ago by Kammerling-Onnes,
superconductivity has remained one of the most active field of
research in physics. The study of this dramatic feature of solids
at low temperature has been the source of countless new concepts
and applications, from the discovery of its microscopic origin by
Bardeen, Cooper and Schrieffer (BCS), to the development of MRI in medical imaging or the invention of SQUIDs
(Superconducting Quantum Interference Devices) which constitute
the most precise magnetic field probes. Nowadays, they are used in
the development of quantum computing for the experimental
realization of q-bits, and the understanding of
the microscopic origin of high critical temperature
superconductivity remains one of the most famous open problem in
physics.

Recently, with the observation of superfluidity in ultra-cold
Fermi gases, a promising connexion has been made between atomic
and solid state physics \cite{inguscio2006ultracold,bloch2008many}. In particular, the experimental
exploration of the phase diagram of atomic fermionic superfluids
as a function of the strength of attractive interactions has
confirmed the existence of a smooth crossover between the BCS
regime of weakly attractive fermions and the Bose-Einstein
condensation (BEC) of deeply bound pairs proposed in the early 80's by
Leggett~\cite{Leggett81}, Nozi\`eres and Schmitt-Rink~\cite{nozieres1985bose}.
Another intriguing issue
recently addressed experimentally is the fate of a superconductor
when the spin populations are imbalanced. Indeed, in the standard
BCS theory, superconductivity arises from Cooper pairing of
opposite spin fermions, and is therefore sensitive to a
mismatch between the Fermi surfaces of the two spin species.
 This fundamental question has in fact been raised in various branches of physics,
including exotic superconductivity~\cite{matsuda2007}
in heavy fermions, organic compounds or cuprates, and nuclear physics~\cite{alford2008},
each field bringing its own peculiarities to the topic.

Mean field theoretical foundations on this issue were laid in the 60's but the phase diagram of a polarized Fermi gas remained
unexplored experimentally until recent work in Rice University, MIT
and ENS on ultra-cold Fermi gases. In what follows, we
present a brief account of the experimental findings made by these
groups. We will show how they triggered new theoretical ideas
and led to a more refined understanding of these
systems.  Note that this review voluntarily focuses on aspects closely related to experiments on ultra-cold Fermi atoms and leaves out more speculative issues. More information can be found in related reviews, in particular \cite{inguscio2006ultracold,bloch2008many,giorgini2008theory,sheehy2010imbalanced}.

\section{The founding fathers}

\begin{figure}
\centerline{\includegraphics[width=0.7\columnwidth]{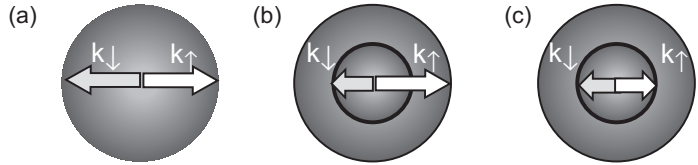}}
\caption{Pairing in an imbalanced Fermi gas. a) In the classical BCS scenario,
particles at the Fermi surface form zero center of mass momentum
Cooper pairs. When the density of spin up and spin down particles are
different, the two Fermi surfaces are mismatched and this mechanism
must be modified. (b) FFLO: If particles with opposite spins form
pairs at the surface of their respective Fermi seas, the Cooper pairs
acquire a momentum, which can be interpreted as a spatial modulation
of the order parameter. (c) Sarma: forming zero momentum pairs implies
opening a gap inside the majority Fermi sea.}
\label{pairingform}
\end{figure}

A significant progress towards the understanding of superconductivity
was achieved when Cooper~\cite{cooper1956} realized that an arbitrarily
weak interaction
pairs electrons with opposite spins in the presence of a Fermi sea.
Bardeen, Cooper and Schrieffer then proposed a variational form for the ground state wavefunction consisting of
a coherent superposition of these Cooper pairs~\cite{bardeen1957theory}.
This wavefunction lies at the core of the standard BCS theory and takes the form
\begin{equation}\label{BCS}
| \psi \rangle = \prod_{k} \left( u_k + v_k c_{k\uparrow}^\dagger c_{-k\downarrow}^\dagger \right)\left|0\right\rangle
\end{equation}
where $\left|0\right\rangle$ is the vacuum and $u_k$ and $v_k$ are variational parameters satisfying the constraint $u_k^2 + v_k^2=1$. $c_{k\uparrow}^\dagger c_{-k\downarrow}^\dagger$
creates a Cooper pair with zero total momentum and the ratio $v_k/u_k$ is
the wavefunction of a single Cooper pair in Fourier space (Fig. \ref{pairingform}.a).
In the standard BCS treatment,
the self-consistency of the theory is applied
to the order parameter $\Delta$ which  sets the gap for single-particle excitations.
The complete pairing of spin $\uparrow$ electrons with spin $\downarrow$
electrons in the BCS wavefunction~(\ref{BCS}) requires the spin populations
to be exactly balanced. Quite naturally, the case of unbalanced population
was first addressed a few years only after the BCS proposal. The question
is of fundamental interest because the spin polarization opposes to the BCS
pairing and eventually destroys superconductivity.
Clogston~\cite{clogston1962ulc} and Chandrasekhar~\cite{chandrasekhar1962} independently
considered the case where the spin
polarization originates from the Zeeman coupling of the electron spin
to an external magnetic field $B$.
They had in mind a situation where the Meissner effect is absent:
the field penetrates the complete metallic sample
and the orbital coupling to the vector potential is negligible. This paramagnetic
limit is characterized by the chemical potential difference $\bar{\mu} = (\mu_\uparrow
- \mu_\downarrow)/2 = \gamma B/2$ between the two spin species. Here $\gamma$ denotes
the gyromagnetic factor for conduction electrons.
Clogston and Chandrasekhar  found a first order transition at $\bar{\mu} = \Delta_0/\sqrt{2}$
above which superconductivity disappears and the normal state takes over the BCS
ground state. $\Delta_0$ is the zero temperature gap for $\bar{\mu}=0$.
This result is simply obtained from the comparison of Gibbs energies, $E = E_0 - N_0 \Delta_0^2/2$
for the {\em unpolarized} BCS state~(\ref{BCS}), and $E =  E_0 - N_0 \bar{\mu}^2$ for the {\em partially polarized} normal state.
$E_0$ is the energy of the normal state in the balanced case and $N_0$ the density of states
at the Fermi level. In fact, protected by the energy gap, the BCS state~(\ref{BCS})
is not modified when $\bar{\mu} \ne 0$.
In particular, spin populations remain equal and the order parameter $\Delta$ remains equal to $\Delta_0$.
The first order nature of the transition is thus a consequence
of the inability of the BCS state to react to a polarization constraint. The energy
for flipping a spin down electron varies like $\Delta_0 - \bar{\mu}$ with the chemical
potential difference $\bar{\mu}$, but remains nevertheless positive at the transition.

Unfortunately,
for superconductors the Clogston-Chandrasekhar (CC), or Pauli, limit corresponds in practice to very high magnetic fields.
In most metallic compounds, superconductivity is limited by orbital pair-breaking
effects and the corresponding upper critical field is much smaller than the Pauli limit.
Orbital effects also lead to a different physics with the emergence of a vortex lattice below the
upper critical field. Pauli limited superconductors are rare and
therefore correspond to specific conditions
where the orbital coupling is suppressed, for instance 2D layers with an in-plane field.
For more detail, we refer the reader to the concise but exhaustive review~\cite{matsuda2007}
where the competition between Zeeman and orbital coupling is discussed in type-II superconductors.

The details of the CC first order transition were later investigated by
Sarma~\cite{sarma1963}. For $\bar{\mu} \ge \Delta_0/2$, the normal state is locally energetically stable and an additional
unstable solution of the gap equation appears.
This intermediate unstable phase, often called the Sarma (or breached-pair~\cite{gubankova2003},  Fig. \ref{pairingform}.b) phase
in the literature, connects the normal state when $\bar{\mu} = \Delta_0/2$ to the fully paired
BCS state $\Delta = \Delta_0$ when $\bar{\mu} = \Delta_0$. In the parameter range $\Delta_0/2
\le \bar{\mu} \le \Delta_0$, the variational energy as a function of $\Delta$ exhibits
the typical features of a first order transition in the framework of Landau theory: two
local minima located at $\Delta = 0$ and $\Delta_0$, {\it i.e.} the normal and BCS phases,
 separated by a local maximum (the Sarma phase)
at $\Delta^2 = \Delta_0 ( 2 \bar{\mu} - \Delta_0)$.
The Sarma phase shows interesting properties~\cite{liu2003igs,shovkovy2003} that
have raised some interest~\cite{wu2003,he2006loff,gubbels2006sarma,dao2008,chien2007superfluid}.
Since $\Delta < \bar{\mu}$ in this phase, a significant number of Cooper pairs can be broken
and the system becomes polarized in contrast with the standard BCS
state~\cite{Iskin2006}.
 Hence the Sarma phase
possesses at the same time a superfluid component and gapless
fermionic excitations between two Fermi surfaces.
It is nevertheless an unstable phase~\cite{sarma1963,wu2003}.
It has been noted~\cite{gubankova2003} that the Sarma phase is a local energy minimum when the spin populations,
and not the chemical potentials, are imposed. However the phase is then unstable against phase
separation~\cite{bedaque2003,caldas2004}.

Of course, these conclusions strictly hold only in the weak coupling limit.
At unitarity, where the scattering length is infinite, Quantum Monte Carlo calculations have shown~\cite{carlson2005atc} that the gapless (Sarma) phase and the
phase-separated BCS-normal state mixture are nearly degenerate in energy. The Sarma gapless phase was also found~\cite{dao2008}
to be stabilized in the presence of an optical lattice
at intermediate coupling. In any case, no experimental evidence of a Sarma-type phase has been found
so far.

An original idea  was put forward independently by Fulde and Ferrell~\cite{fulde1964}, Larkin and Ovchinnikov~\cite{larkin1964}:
the lack of polarization flexibility of the BCS wavefunction can be improved by giving a finite
momentum to the Cooper pairs (Fig. \ref{pairingform}.c). The resulting order parameter $\Delta ({\bf r})$ is spatially varying.
The finite momentum moves the Fermi surfaces of the two spin species with respect to each other,
bringing them closer in one direction. This effect improves the pairing but costs some kinetic
energy. The displacement of the Fermi surface also leads to pair breaking and allows the system to polarize,
thus relaxing part of the polarization constraint. As expected, the pair breaking is enhanced
in the regions where the order parameter  $\Delta ({\bf r})$ vanishes. There is often some confusion
in the literature
on the definition of the Fulde-Ferrell-Larkin-Ovchinnikov (FFLO) phases. They do not simply reduce
to an helicoidal structure, $\Delta ({\bf r}) = \Delta e^{i {\bf q}\cdot {\bf r}}$, or even
a sinusoidal one, for instance $\Delta ({\bf r}) = \Delta \cos {\bf q}\cdot {\bf r}$. They correspond
generally to an inhomogeneous order parameter with a periodic structure that does not contain vortices.
The characteristic length for the spatial variations is the coherence length of the superconductor.
The phase transition between the normal state and the superfluid FFLO phase was calculated in
Refs.~\cite{mora2004ltf,combescot2005transition,mora2005transition}
using the quasiclassical Eilenberger equations~\cite{eilenberger1968,larkin1965,serene1983}.
This is a first order transition, located at
$\bar{\mu} = 0.781 \Delta_0$ (the corresponding spinodal line is for $\bar{\mu} = 0.754 \Delta_0$),
and it slightly extends the superfluid domain when compared to
the homogeneous BCS state as shown in Fig.~\ref{Fig4.2}. The order parameter has a cubic
structure at the transition and evolves into a square and then a one-dimensional pattern
as the temperarure is increased~\cite{combescot2002}.

Interestingly, the structure of the FFLO phases as one penetrates inside
the superfluid remains an open problem, even at weak coupling. More is known on the transition between the homogeneous
standard BCS phase and the FFLO phase.
In analogy with magnetic systems, the superfluid can form domain walls corresponding to a sign
change of the order parameter over a spatial size on the order of the BCS coherence length. Far from
the wall, the order parameter is constant. In addition to the usual gapped single-particle excitations, the
wall shelters states with energies below the gap~\cite{buzdin2005,yoshida2007}. Filling these states with spin up particles releases part
of the polarization constraint, while the bending of the order
parameter within the domain wall costs some condensation energy.
A second order BCS-FFLO transition~\cite{burkhardt1994,matsuo1998,houzet1999} occurs when the wall energy
becomes negative: a one-dimensional array of walls is formed, starting with infinite periodicity, which alternatively
reverses the sign of the order parameter.
The transition is located~\cite{matsuo1998} at  $\bar{\mu} \simeq 0.666 \Delta_0$.
As the chemical potential difference $\bar{\mu}$ increases, the periodicity decreases merging the order
parameter shape to a sinusoidal form, typical of FFLO phases.
An analytical solution to this domain wall structure was obtained
in the one-dimensional case~\cite{machida1984,buzdin1983,parish2007quasi}.
Additional information on the physics of FFLO phases can be found in
the complementary reviews~\cite{matsuda2007,casalbuoni2004,combescot2007int}.

In the context of cold atoms, the FFLO phases have been extensively
studied theoretically along the BEC-BCS crossover
and especially close to unitarity \cite{sheehy2010imbalanced}. These various studies include
mean-field calculations~\cite{mizushima2004dis,mizushima2007imbalanced,hu2006mean,sheehy2007bec,he2007single}, large N approach~\cite{veillette2008radio} or Density Functional Theory (DFT) \cite{bulgac2008unitary}.
Including the trapping potential beyond the local density
approximation has been argued to be necessary for describing the FFLO
spatial oscillations~\cite{castorina2005nsp,kinnunen2005sif,mizushima2007imbalanced}.
Nevertheless the same oscillations in the order parameter were interpreted in Ref.~\cite{liu2007mean}
to be a finite-size effect.
With or without the trap, none of these approaches is completely exact and
the results are therefore expected to be qualitative.
Most works predict that the FFLO phases exist outside
the weak coupling domain but there is no consensus on
the value of  the interaction strength  for which they disappear.

\section{Polarized Fermi gases}

\subsection{Fermionic superfluidity with ultra-cold gases}


As mentioned earlier, Meissner effect prevents reaching the Pauli limit in most superconductors. As a consequence, the  Clogston-Chandraskhar instability scenario or the existence of Fulde-Ferrell-Larkin-Ovchinnikov phases  were not unambiguously tested experimentally before the observation of superfluidity in ultra cold Fermi gases. These experiments were made possible by the development of techniques of laser cooling and trapping of atoms that led to the observations of the first Bose-Einstein condensates of alkali vapors in 1995 \cite{cornell2002nobel,ketterle2002nobel}. Building on this breakthrough, a new generation of experiments was initiated soon after to cool down fermionic atoms and quantum degenerate Fermi gases were observed for the first time in 1999 by the group of JILA  \cite{demarco1999onset}. Fermionic superfluidity was obtained in 2003 \cite{jochim2003bose,greiner2003emergence,Zwierlein03obe,bourdel2004esb} and allowed for the exploration of BEC-BCS crossover physics, a theoretical scenario  bridging the gap between the Bardeen-Cooper-Schrieffer mean-field theory describing the behavior of weakly attracting fermions (scattering length $a$ small and negative), and the strongly attractive regime ($a$ small and positive) where the system behaves as a Bose-Einstein condensate of tightly bound dimers. Remarkably, although the theoretical foundations of this crossover physics had been laid in the early eighties by the pioneering works of Nozi\`eres, Schmitt-Rink  and Leggett \cite{Leggett81,nozieres1985bose}, its first experimental confirmation was only made possible by the possibility of tuning interactions in ultra-cold atom vapors using Fano-Feshbach resonances. Another asset of cold atoms against classical condensed matter systems is the long spin relaxation time which offers the possibility of  controlling spin populations using radio-frequency fields or optical pumping and keeping spin imbalances for long times. In addition to Feshbach resonances, several unique investigations tools were also developed  and in the following we briefly review the main progresses that were achieved in the first years following the observation of fermionic superfluidity (note that we focus here on the tools used latter on for the study of spin imbalanced gases and as a consequence, we leave out important experiments such as the pair projection method \cite{Regal2004Observation,Zwierlein2004Condensation} or the study of collective modes \cite{bartenstein2004cmo} and let the reader refer to the more exhaustive review \cite{inguscio2006ultracold}).

\subsubsection{Scattering length and Feshbach resonances}
Although ultra-cold gases are highly dilute, interactions still play a crucial role in their properties. Nevertheless, the description of interactions can be simplified by Pauli principle and the low temperature of these systems. Indeed, at $\sim 1$~$\mu$K, the thermal wavelength of an atomic vapor is $\sim 0.1$~$\mu$K, and is therefore much larger than the typical range of interatomic potentials ($\sim 1$~nm). This means that matter waves do not resolve the exact details of the potential, which can therefore be described as a zero-range contact potential. In 3D, a Dirac potential is highly singular and leads to divergences which need to be regularized. Several strategies can be followed, for instance the use of a non-local pseudo potential defined by

$$V_{\rm pseudo}(r)=g\partial_r(r\cdot)\delta(\bm r),$$

\noindent where the coupling constant is usually written as $g=4\pi\hbar^2 a/m$ and $a$ is the scattering length \cite{inguscio2006ultracold}. In the case of a two-body problem, this pseudo-potential is characterized by the following properties:

The scattering amplitude for a relative momentum $k$ is given by $f(\bm k)=-a/(1+ika)$. We see that the interpretation of $a$ is here straightforward, since it corresponds to the low energy limit of the scattering amplitude, and is thus associated with the scattering cross-section. An important point is the limit $|a|=\infty$, where the scattering length diverges, but the scattering amplitude stays finite and reaches an universal value $1/ik$. It can be demonstrated that this maximum value is imposed by the unitarity of the $S$-matrix describing scattering events, and this regime is therefore called the {\em unitary limit}.

For $a>0$, the two body-potential possess a bound state of energy $E_b=-\hbar^2/ma^2$. We see here that the unitary limit is associated with the disappearance of the two-body bound state, a generic feature which can be interpreted in terms of scattering resonances between incoming free particles and low-lying bound states. This connexion between the sign of the scattering length and the properties of weakly bound states suggests that the regime where $a$ is small and positive corresponds to a strongly attractive potential, while on the contrary, the $a$ negative and small regionn corresponds to a weakly attractive regime where the interatomic potential is too shallow to overcome quantum fluctuations and maintain a bound state.

\begin{figure}
\centerline{\includegraphics[width=\columnwidth]{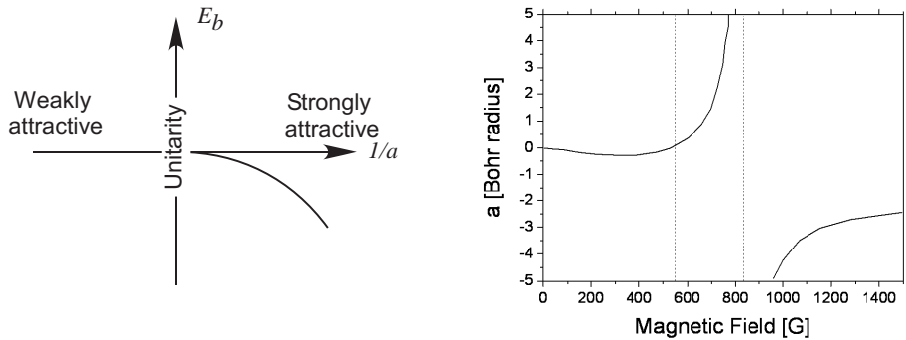}}
\caption{Left: At low energy, the short range interatomic potential is characterized by the scattering length $a$. A weakly bound state of energy $E_b=-\hbar^2/ma^2$ is stable for $a>0$. $|a|=\infty$ is associated with a universal regime known as the unitary limit. Right: the scattering length can be tuned using a magnetic field (here scattering length of $^6$Li between the two spin states $|F=1/2,m_F=\pm1/2\rangle$ of the ground state hyperfine manifold). The divergences are known as Feshbach resonances and are associated with the resonance between a weakly bound state and scattering states. Fermions being stable close to a Feshbach resonance, it is possible to study the crossover between strongly and weakly attractive regimes. }
\label{Fig3.4}
\end{figure}

In cold atoms, the scattering length can be be tuned by imposing an external magnetic field tuning the position of the bound state, the so-called Feshbach resonance \cite{Tiesinga1992Conditions,Tiesinga1993Threshold,Moerdijk1995Resonances,Vogels1997Prediction}. Contrarily to bosonic system which are unstable when $a$ diverges \cite{Courteille1998Observation,Roberts1998Resonant}, it could be demonstrated both theoretically \cite{petrov2004weakly} and experimentally \cite{Cubizolles2003Production,jochim2003pure,regal2004lifetime} that Pauli principle was actually suppressing inelastic collisions in the regime of large scattering lengths. This unique property allows one to probe on a single atomic system the full crossover between strongly and weakly attractive domains (Fig. \ref{Fig3.4}).

\subsubsection{Density profile and equation of state.}
\label{SectionDensityProfile}

Due to the external optical or magnetic confinement, the density
profile of an ultra-cold gas is inhomogeneous and complicates
comparison with theories usually developed for homogeneous
systems. However, in most cases, the size of the system is large
enough to define a mesoscopic length scale where the properties of the
cloud can be described locally by the equation of state of an
homogeneous gas.
This is the Local Density Approximation (LDA). The local chemical
potential $\mu_\sigma(\bm r)$ can be defined, as the energy necessary to add
a particle of spin $\sigma$ at position ${\bm r}$. The chemical potential  $\mu_{\rm hom}[n_\uparrow,n_\downarrow]$ of an
homogeneous gas with the density $n_\sigma$ is added to the trapping
potential $U_\sigma$ such that $\mu_\sigma(\bm r) = \mu_{\rm hom}[n_\uparrow(\bm r),n_\downarrow(\bm r)]+U_\sigma(\bm
r)$. At equilibrium, the local chemical potential $\mu_\sigma(\bm r)$ is
homogeneous and the density profile can be calculated from the
resolution of the equation $\mu_{\rm hom}[n_\uparrow(\bm r),n_\downarrow(\bm r)]+U_\sigma(\bm r)=\mu_0$, where $\mu_0$ is a constant and is the global chemical potential of the system.

Let us consider the special case of a spin balanced fermionic superfluid at zero temperature, with $\mu_\uparrow=\mu_\downarrow=\mu$. Dimensional analysis shows that the equation of state $\mu_{\rm hom}(n)$  takes the form

\be
\mu_{\rm hom}(n)=E_F f(1/k_Fa),
\label{Eqn2.3}
\ee

\noindent where $k_F=(6\pi^2 n)^{1/3}$ is the Fermi wave-vector of an ideal spin-polarized Fermi gas of density $n$, $E_F=\hbar^2 k_F^2/2m$ is the associated Fermi energy and $f$ is some dimensionless function that must be determined from a microscopic theory and interpolates between the BEC and BCS sectors of the crossover. At unitarity ($|a|=\infty$), the macroscopic equation (\ref{Eqn2.3}) simplifies greatly since we have

\be
\mu_{\rm hom}=\xi E_F,
\label{Eqn2.2}
\ee

\noindent  where $\xi=f(0)$ is just some numerical factor \cite{Heiselberg01fsl}. Remarkably, we observe that, up to this numerical  factor, this is exactly the equation of state of an ideal gas: in other words, although we are considering the unitary regime, where interactions induce strong quantum correlations at the microscopic level, the macroscopic static properties are as simple as that of an ideal gas. In particular, in a harmonic trap the density profile is readily obtained  using LDA, and the radius of the cloud is simply $R=R_{\rm NIFG}\sqrt[4]{\xi}$, where $R_{\rm NIFG}$ is the radius of a non interacting Fermi gas with the same atom number and in the same trapping potential. This relationship suggests a straightforward way to determine the interaction parameter $\xi$, by simply measuring the ratio $R/R_{\rm NIFG}$. This very scheme and some variations have been applied experimentally {\rm in situ} or after time of flight and theory \cite{astrakharchik2004eq,Carlson2003Superfluid,carlson2005atc,Perali2004Quantitative,haussmann2007thermodynamics} and experiments \cite{Bartenstein2004Crossover,kinast2005heat,partridge2006pap,stewart2006potential} now converge towards the value $\xi = 0.42 (1)$.

However, the simple scaling law presented above is valid only at
unitarity and comparison between theory and experiments in the
crossover is more involved. Indeed, atomic density profiles are
obtained in practice by absorption imaging which gives access to an
integrated two-dimensional density $\bar n_{2D}(x,y)= \int dz \, n(x,y,z)$, where the line of sight is taken along $z$. To obtain insight on the local properties of the cloud, it is thus necessary to deconvolve this integration. In the case of a cylindrically symmetric trap, this can be achieved by using a data processing technique inspired from medical imaging technology based on a mathematical operation called Abel transform \cite{Epstein2007Introduction} and applied recently in the case of imbalanced Fermi gases \cite{shin2008des}. An alternate scheme to measure the grand canonical equation of state of the homogeneous gas was proposed recently~\cite{ho2009opdtq}. Indeed, let us consider the double integrated density of spin $i$ atoms defined by $\bar n_i(z)=\int dxdy \, n(x,y,z)$. Using Gibbs-Duhem relation $n_i=\partial_{\mu_i} P$  relating pressure $P$ and density $n_i$, we can write that

\be
\sum_i \bar n_i(z)=\int dxdy \sum_i\frac{\partial P}{\partial\mu_i}.
\ee

Due to the spatial dependence of the chemical potentials, LDA allows one to turn the spatial integral into an integral over $\mu_i$ reading

\be
\sum_i \bar n_i(z)=\int \sum_i \frac{\pi d\mu_i}{m\omega_r^2} \frac{\partial P}{\partial\mu_i},
\ee

\noindent where $\omega_r^2=\omega_x\omega_y$ is the transverse trapping frequency. Integration is straightforward and yields

\be
P(\mu_i(z),T)=\frac{m\omega_r^2}{\pi}\sum_i \bar n_i(z).
\label{Eqn2.1}
\ee

Eq. (\ref{Eqn2.1}) implies that one gets access to the equation of state $P(\mu_i,T)$ of the homogeneous gas simply by measuring the doubly integrated density profile in the trap. This relation is very general and can be  used in a harmonic trap as soon as local-density approximation is valid, for instance in the presence of a spin imbalance or in an optical lattice. Remarkably, the equation of state $P(\mu_i,T)$ contains all the macroscopic information on the system, since using Gibbs-Duhem relation, it is possible to extract missing thermodynamic quantities, such as atom density or entropy. This scheme as been applied for the first time in \cite{nascimbene2009eos} to the case of the finite temperature equation of state of a balanced Fermi gas, but can also be used to probe the properties of imbalanced systems, as we will see later.

\subsubsection{Radio-frequency spectroscopy.}

One of the main features of fermionic superfluidity is the presence of a  pairing gap in the excitation spectrum that constitutes the most direct manifestation of Cooper pairing responsible for the onset of quantum order in these systems. In cold atoms, this quantity can be accessed by spectroscopic tools: for instance, it is possible excite atomic spin degrees of freedom using radio-frequency field, and measure the shift of the resonance induced by interatomic interactions. Several groups have measured the binding energy of fermion pairs in ultra-cold Fermi systems using this scheme \cite{chin2004observation}, starting from pioneering experiments by the groups of JILA and MIT on quantum degenerate Fermi gases \cite{gupta2003radio,Regal2003Creation}, to the recent measurement of the spectral function of a fermionic superfluid at JILA \cite{stewart2008using}. Nevertheless, the interpretation of experimental data is non-trivial since interactions between the superfluid and the final spin state can modify strongly the response of the system. On the theory side, intense effort has been devoted to the understanding of the role of interaction in rf spectroscopy \cite{basu2008final,veillette2008radio,massignan2008twin,punk2007theory,perali2008competition}, and in experiments, it has been pointed out that initial and final spin states could be chosen so as to reduce these spurious effects and permitted an unambiguous measurement of the excitation gap of the system as demonstrated in \cite{schunck2008dfp}.

\subsubsection{Vortices and superfluidity.}

Although in BCS theory pairing and superfluidity arise at the same time, this property is not general. In most cases the two phenomena are decoupled: for instance, in the case of a Bose-Einstein condensate of deeply bound molecules, the formation of the dimers takes place at a temperature much higher than the condensation critical temperature. In practice, the most convincing probe of superfluidity in ultra-cold Fermi gases was obtained by the group of MIT by stirring the cloud using a far detuned laser beam and observing the formation a triangular array of quantized vortices that can be considered as a the smoking gun for the existence of a complex order parameter characterizing the superfluid phase \cite{zwierlein2005vortices}.

\subsection{Ultra-cold Fermi gases with imbalanced spin populations}

Once balanced fermionic superfluidity is achieved with ultracold systems, the experimental study of spin polarized Fermi systems is straightforward. The groups of Rice University and MIT prepared simultaneously  for the first time ultra-cold Fermi gases with imbalanced spin populations \cite{partridge2006pap,zwierlein2006fsi}, and, despite some discrepancies between their results, the two experiments support Clogston-Chandrasekhar's original proposal that the superfluid phase is robust against spin polarization. Indeed, both groups observed that, in the presence of an imbalance in the spin populations, the cloud phase separates along a shell structure. Using Abel reconstruction they observe that a fully paired core with equal spin densities is surrounded by a ``magnetized" rim with imbalanced spin populations \cite{shin2006observation,partridge2006deformation}. By rotating the cloud, it was in addition observed that only this central core could sustain quantized vortices, indicating that the central phase is superfluid and that the magnetized phase is not \cite{zwierlein2006fsi}. This very simple observation constitutes a crystal clear confirmation of the Clogston-Chandrasekhar hypothesis. Indeed, for cold atoms, a polarizing magnetic field is equivalent to a chemical potential mismatch between the two spin species. In the case where the trapping potential is the same for the two spin states (which is the situation realized experimentally), LDA shows that the chemical potential difference $\bar\mu=(\mu_\uparrow-\mu_\downarrow)/2$ is constant in the cloud. Since the gap $\Delta$ decreases with density, the CC scenario predicts that an unpolarized superfluid should occupy the center of the trap, as long as $\Delta$ stays roughly larger than $\bar\mu$, which is qualitatively in agreement with experimental findings.

\begin{figure}
\centerline{\includegraphics[width=\columnwidth]{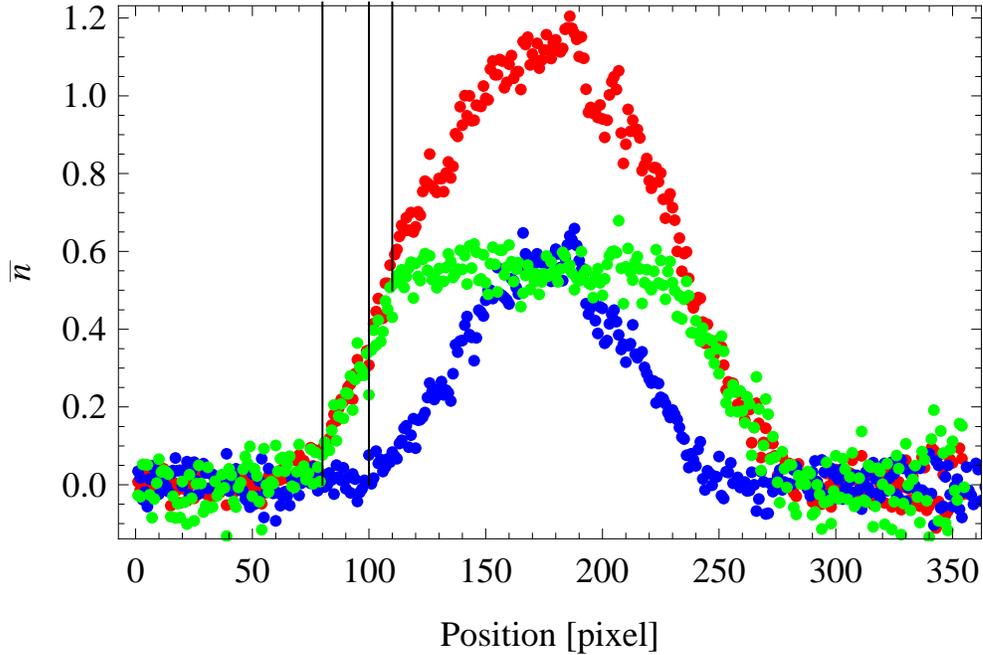}}
\caption{Density profile of a unitary spin imbalanced Fermi gas. From top to bottom: $\bar n_\uparrow$ (majority, red), $\bar n_d=\bar n_\uparrow-\bar n_\downarrow$ (green) and $\bar n_\downarrow$ (minority, blue). At low temperature, the density profile presents a shell structure, with a fully paired superfluid at center, surrounded by a partially polarized normal phase, and at the outside a fully polarized ideal Fermi gas of majority particles. Note that the flat profile of the integrated density difference $\bar n_d$ is the signature of a fully paired core \cite{desilva2006stu}.}
\label{Fig2.1}
\end{figure}

Despite the important similarities in the structure of the central core obtained by the Rice and MIT experiments, major discrepancies exist on the structure of the normal phase at the rim of the trap. Indeed, while Rice's group observes a single fully polarized phase where only majority atoms are present, MIT obtains a richer phase diagram, with a normal component divided between a fully polarized region similar to the one observed at Rice, and a mixed phase where the two spin component are present (Fig. \ref{Fig2.1}). A first consequence of this difference is that apart from the extreme polarization case $N_\downarrow=0$, a superfluid core is always present in Rice's observations. This particular feature contradicts MIT measurements, which observes that above a polarization $P_c=(N_\uparrow-N_\downarrow)/(N_\uparrow+N_\downarrow)$ ($P_c\sim 0.75$ at unitarity, confirmed by recent ENS measurements), the core is no longer superfluid, and only the two normal phases are observed.

\section{Zero temperature phase diagram}

In this section we present theoretical interpretations of the experimental results outlined above. Since most experiments were performed at unitarity, we will focus first on this regime, and then we will later provide an extension to the whole crossover. A key question we wish to answer is whether  a stable partially polarized normal phase exists. This issue constitutes the main difference between the experiments of Rice and MIT and we shall show that this question can be answered by the study of a simpler problem, namely the calculation of the energy for an impurity immersed in a Fermi sea. Although rather academic at first glance, the resolution of this problem will constitute the backbone for the description of the normal phase in MIT's experiment.

\subsection{Scaling law and unitary phase diagram.}

Due to the presence of a trap in experiments, it is more convenient to describe the behavior of the cloud in the grand canonical ensemble, where the thermodynamic properties of the system are fully characterized by the equation of state $P(\mu_{i=\uparrow,\downarrow},T)$ \cite{callen1985thermodynamics}. Using the same dimensional argument as for the balanced unitary gas, this equation of state at unitarity and $T=0$ can be recast as a function of a single dimensionless function $h$ such that

$$\frac{P(\mu_\uparrow,\mu_\downarrow)}{P_0(\mu_\uparrow)}=h(\eta=\mu_\downarrow/\mu_\uparrow),$$

\noindent where $P_0$ is the pressure of a single component ideal gas \cite{chevy2006upa,bulgac07ztt}.

\begin{figure}
   \begin{minipage}[c]{0.46\linewidth}
      \includegraphics[width=\linewidth]{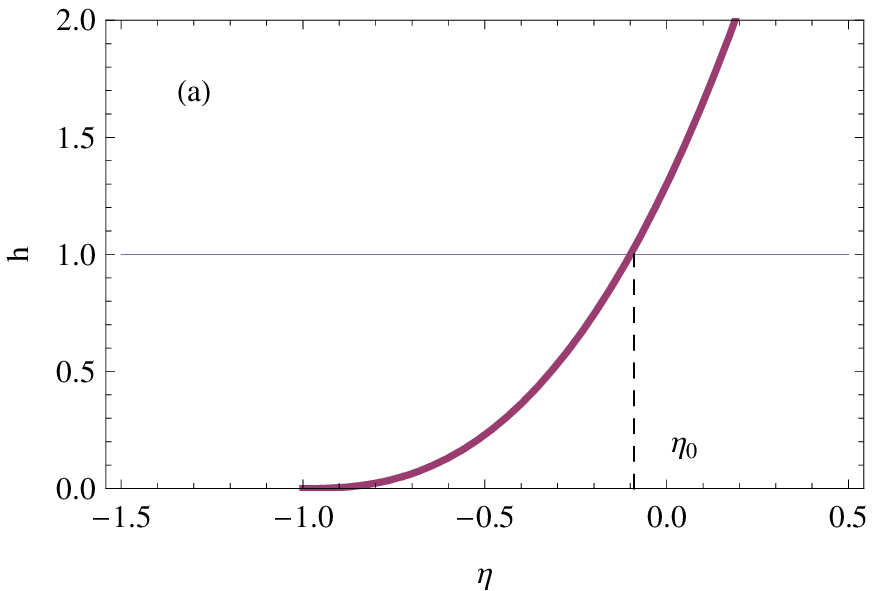}
   \end{minipage} \hfill
   \begin{minipage}[c]{.46\linewidth}
      \includegraphics[width=\linewidth]{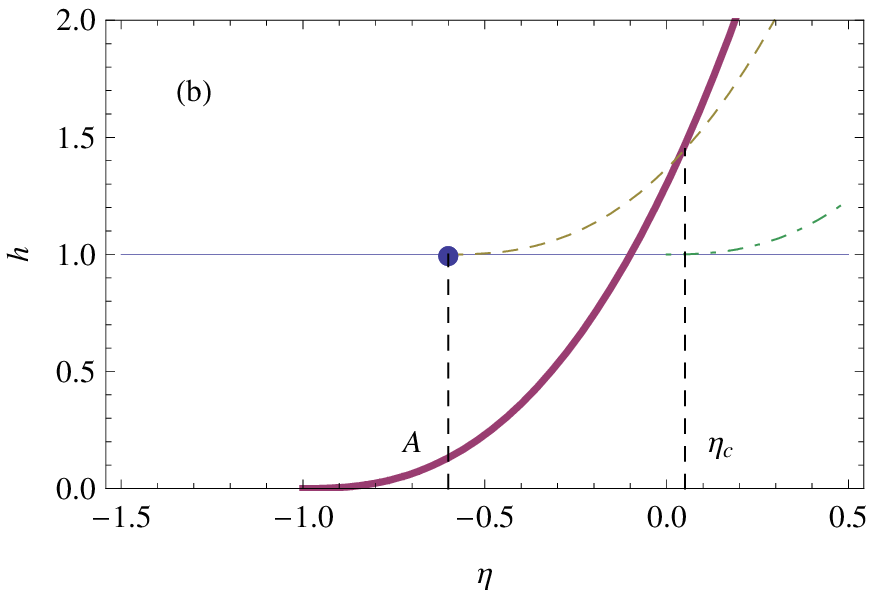}
   \end{minipage}

\caption{Sketch of the equation of state of an imbalanced Fermi gas. From the thermodynamics relation $\Omega=-PV$, the stable phase is that with the highest $h$. (a) Rice experiment. Only two phases were observed: a fully polarized ideal gaz (black horizontal line) and a fully paired superfluid (red line). They cross at $\eta_0\sim -0.1$ which indicates the position of the Clogston-Chandraseckar limit in this situation.  (b) MIT experiment. The dashed and dot-dashed line give a sketch of stable and instable intermediate normal phases. Here $\eta_c$ is the normal-superfluid threshold, and $A$ the boundary between the partially and fully polarized phases. Graphically, we observe that the stability criterion for the intermediate phase is $A<\eta_0$. Moreover,  $A\mu_\uparrow$ can be interpreted physically as the energy of the last minority atom removed from the majority Fermi sea, and is thus closely related to the Fermi-polaron problem.  }
\label{Fig3.1}
\end{figure}

 A sketch of $h$ corresponding to Rice's observations is displayed in  Fig.~\ref{Fig3.1}.a. On the one hand, the equation of state in the normal phase is that of an ideal gas of spin up particles, hence $h=1$. On the other hand,  the knowledge of the equation of state of a balanced superfluid, Eq.~(\ref{Eqn2.2}), yields in the superfluid core $h=(2\xi)^{-3/2}(1+\eta)^{5/2}$ \cite{chevy2006dpt}. In this diagram, the CC limit corresponds to $\eta_0=(2\xi)^{3/5}-1\sim -0.1 $ where the two curves meet. Indeed, since pressure is related to the grand potential $\Omega$ and the volume $V$ by the relation $\Omega=-PV$, the most stable phase at given chemical potentials is the one with the highest pressure. For $\eta>\eta_0$ the most stable phase is the fully paired superfluid, while  the fully polarized ideal gas is favored for $\eta<\eta_0$.

Let us now consider what would be the phase diagram corresponding to MIT's observations, as shown in Fig.~\ref{Fig3.1}.b.
The intermediate mixed phase is a partially polarized normal phase that intersects the fully polarized Fermi gas at $\eta = A$.
 Physically, the corresponding phase transition is interpreted as the removal of the last spin down particle from the partially polarized normal phase.
Therefore the dimensionless parameter $A$ is determined from the energy $E$ of a single minority atom immersed in a Fermi sea of majority particles. By definition, this energy is the chemical potential $\mu_\downarrow$, while $\mu_\uparrow$ is the Fermi energy $E_F$ of the spin up Fermi sea, hence $A=E/E_F$ \cite{bulgac07ztt,chevy2006upa,lobo2006nsp}.
The value of $A$ is crucial to determine the relevant scenario for the density profiles. For $A > \eta_0$, the mixed phase is unstable (dot-dashed line  in Fig.~\ref{Fig3.1}.b),
always preempted by the superfluid phase, and  RICE's features are recovered.
However, for $A < \eta_0$, the mixed phase  (dashed line) extends between $\eta = A$ and $\eta = \eta_c$, in agreement with MIT's observations.

\subsection{The Fermi polaron.}

In the previous section, we have seen that the stability of a partially polarized phase could be related to the physics of an impurity immersed in a Fermi sea, the so-called Fermi-polaron named after the solid state physics polaron that describes the interaction of an electron with a bath of (bosonic) phonons. Despite its simplicity (one impurity immersed in a non interacting Fermi sea), the lack of a small parameter in this problem
makes it strongly interacting and  essentially non-perturbative. Monte-Carlo methods are {\em a priori} required to calculate the energy spectrum of the impurity \cite{lobo2006nsp,prokof'ev08fpb}.

However, to demonstrate the stability of the normal phase observed at MIT, we just need an upper bound to this energy. A variational method is thus sufficient to settle the issue at stake. In the polaron picture, an impurity immersed in some medium forms a quasi-particle composed of the bare impurity dressed by excitations of the surrounding environment. In the present case, elementary excitations of a Fermi sea are particle-hole pairs that shroud the spin down atom to form the Fermi-polaron. Since the impurity is alone, we can assume that the effect of the impurity on the Fermi sea remains weak, and we can try to work in a subspace where a single particle-hole pair is created by the presence of the spin down atom. Variational equations can be worked out analytically, leading  for small $\bm p$ to a dispersion relation

$$E_{\bm k}=AE_F+\frac{\hbar^2k^2}{2m^*}+...,$$

\noindent analog to that of a free particle with renormalized parameters $A\sim -0.6$ and $m^*=1.17 m$ for $a=\infty$ \cite{chevy2006upa,combescot2007nsh}. Let us stress that  this result is variational, and gives us access to an upper bound of the energy. The true value of $A$ is then strictly speaking smaller that -0.6. Nevertheless, this bound is sufficient to conclude that $A$ is indeed smaller than $\eta_0$ and to demonstrates rigorously the existence of a partially polarized phase between the fully polarized and fully paired sectors of the phase diagram.

Even more remarkably, it appears that the variational values found here are strikingly close to the actual predictions  $A=-0.59(1)$ and $m^*=1.09m$ for the variational Fixed Node Monte-Carlo~\cite{lobo2006nsp} and $A=-0.61(1)$ and $m^*=1.20(1)m$ for diagrammatic Monte-Carlo~\cite{prokof'ev08fpb}. This coincidence can be explained by the relatively weak probability of excitation of a particle-hole pair. Indeed, in the variational calculation this probability is only $\sim 25$\%, despite the strength of interactions at unitarity . This is also confirmed by the moderate modification of the polaron mass which is close to its bare value, and also by a systematic expansion of the polaron energy as a function of the number of particle hole-pairs excited which demonstrated that the associated series in $k_F$ was converging quite fast \cite{prokof'ev08fpb,combescot2008nsh}.

Let us now discuss the equation of state of this intermediate normal
phase. As first proposed in \cite{lobo2006nsp}, if one assumes that
the impurities keep their fermionic nature, a natural description of
their collective behavior at low density of spin down particles is
that of an ideal Fermi gas of polarons, a picture confirmed by Monte
Carlo simulations of the equation of state. In this regime, it was
shown in Ref.~\cite{nascimbene2009eos,mora2010Normal} that
the pressure of the normal mixture is the sum of the pressures
of an ideal gas of majority atoms and an ideal gas of polarons, i.e.
\be
P=\frac{1}{15\pi^2}\left(\frac{2m}{\hbar^2}\right)^{3/2}\mu_\uparrow^{5/2}
+\frac{1}{15\pi^2}\left(\frac{2m^*}{\hbar^2}\right)^{3/2}\left(\mu_\downarrow-A\mu_\uparrow\right)^{5/2}.
\label{Eqn3.2}
\ee
Despite its seemingly non-interacting form, Eq.~(\ref{Eqn3.2}) does
include interactions between polarons as a result of the
$\mu_\uparrow$ dependence in the second term.  Translating
Eq.~(\ref{Eqn3.2}) into the canonical ensemble, an exact relation
between the dominant polaronic interaction and the parameter $A$ was
derived in Ref.~\cite{mora2010Normal}, in relatively good agreement with
fixed node Monte-Carlo simulations (within a few percents).

We observe that the equation of state of the normal phase meets that of the superfluid at a value $\eta_c\sim 0.065$. It corresponds to the maximum chemical potential mismatch that the superfluid can sustain before turning into the normal state. This determination of the Clogston-Chandreskhar limit at unitarity allows one to reconstruct the density profile in a trap, as well as calculate the critical polarization at which the superfluid core vanishes in MIT's experiments \cite{recati2008ris}, yielding $P_c\sim 0.77$, in close agreement with the value measured at MIT and ENS.

\subsection{Extension to the crossover and the polaron-molecule transition.}

\begin{figure}
   \begin{minipage}[c]{0.46\linewidth}
      \includegraphics[width=\linewidth]{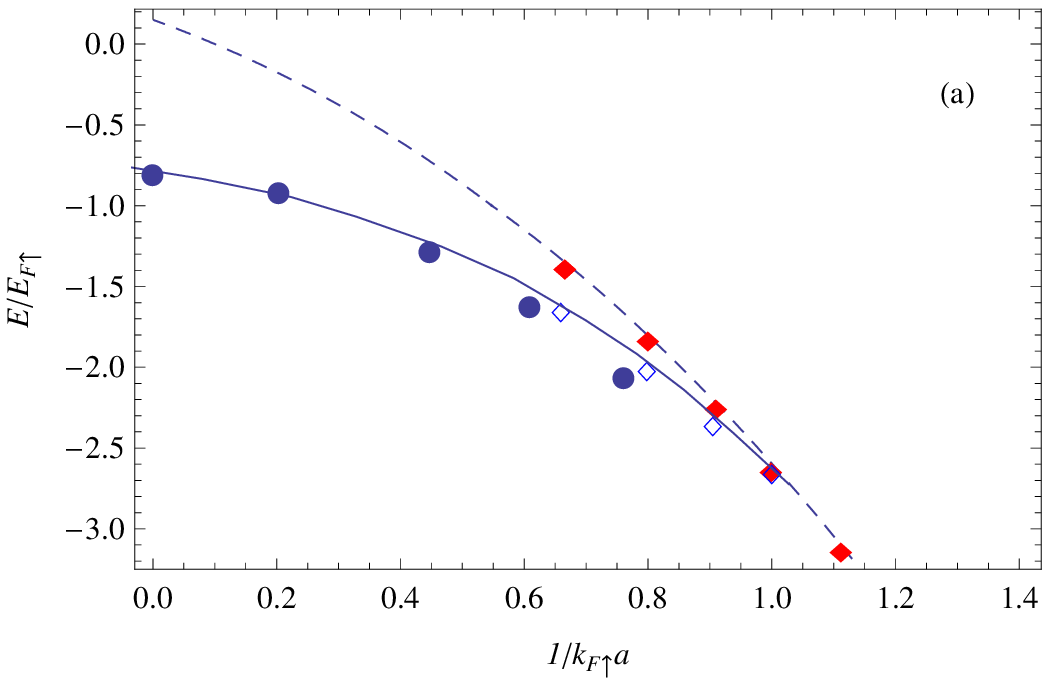}
   \end{minipage} \hfill
   \begin{minipage}[c]{.46\linewidth}
      \includegraphics[width=\linewidth]{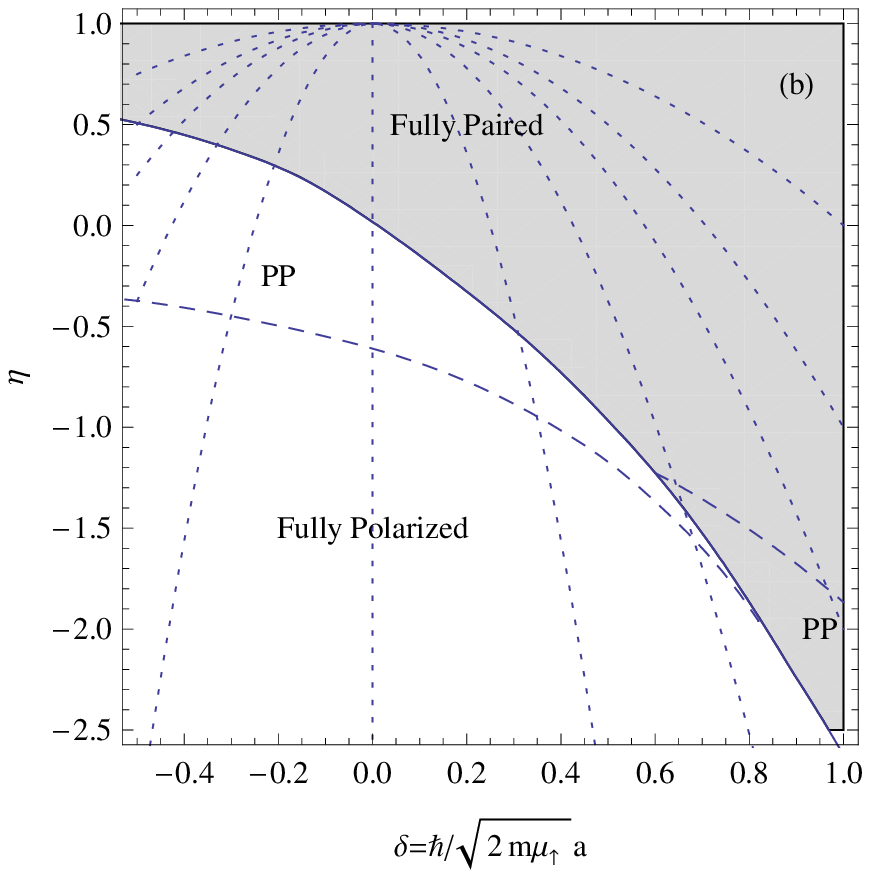}
   \end{minipage}
\caption{Imbalanced fermi gas in the BEC-BCS crossover. (a) energy of an impurity immersed in a Fermi sea and molecule/polaron transition. Solid (dashed) line, variational calculation of the polaron (molecule) \cite{punk2009polaron,mora2009ground,combescot2009analytical}. Solid circles: experimental results from MIT \cite{schirotzek2009ofp}. solid (empty) diamonds: Diagrammatic Monte-Carlo calculation of molecule (polaron) \cite{prokof'ev08fpb}. (b) phase diagram in the crossover in the grand canonical ensemble. The gray region indicates the superfluid sector. PP stands for partially polarized. Solid (resp. dashed) lines correspond respectively to first and second order transitions. Note that beyond $\delta=1.7$, the normal/superfluid transition becomes second order. The dotted lines correspond to the ``path" followed in a trap.}
\label{Fig3.2}
\end{figure}

In the previous section, we have shown that the normal phase at
unitarity could be  described quantitatively by a Fermi liquid theory
involving Fermi polarons whose physical properties could be obtained
with a rather good accuracy from a simple variational
calculation. This picture is still valid in the BCS sector of the
crossover, since by construction, the variational ansatz used to
describe the Fermi-polaron coincides with second order perturbation
theory in the limit $a\rightarrow 0^-$. In the BEC sector, however,
the Fermi liquid scenario must break down at some point since in the
far BEC limit, we expect the impurity to form a tightly bound dimer
with one majority atom, giving rise to a bosonic, rather than
fermionic, behavior of the dressed quasi-particle. This
molecule/polaron transition was unveiled by Monte-Carlo simulation of
the impurity problem \cite{prokof'ev08fpb}. This calculation shows
that the transition takes place at $1/k_{F\uparrow} a = 0.91(2)$, and
that above this threshold the energy of the polaron  is accurately
described by that of a point-like boson interacting with the Fermi sea
by a mean-field energy $g_{\rm ad} n_\uparrow$, where the atom-dimer
coupling constant $g_{\rm ad}$ is characterized by the well known
atom-dimer scattering length $a_{\rm ad} \sim 1.18 a$
\cite{skorniakov1957three}. This unexpected robustness in the validity
of the mean-field approximation was  explained later on by an
extension of the variational description of the polaron to the
molecular sector
\cite{mora2009ground,punk2009polaron,combescot2009analytical}.
Similarly to the BCS/unitary case, the quasi-particle is described as
a dimer dressed by single particle-hole pair of the majority Fermi
sea. Due to the composite nature of the dimer, the calculations are a
little more involved than in the fermionic case, but can be reduced to
a single integral equation that is solved numerically. The energy
displayed in Fig. \ref{Fig3.2}.a demonstrates a remarkably good agreement between Monte-Carlo simulations, variational calculations and rf experiments discussed below.

From bosonic to fermionic along the crossover, the change in the nature of the impurity immersed in the Fermi sea of majority atoms has strong consequences on the structure of the phase diagram. Indeed, while no qualitative difference is expected between unitarity and BCS sectors (essentially the same three phases that we discussed above), the BEC regions is dominated by a completely different physics \cite{pilati2008psp}. As suggested by the molecular scenario, we can describe the system in the BEC limit as a mixture of a Bose-Einstein condensate of dimers and a Fermi sea of excess fermions using the mean-field equation of state \cite{viverit2002ground,Leyronas2009Equation}

$$\frac{E}{V}(n_F,n_B)=-\frac{\hbar^2}{ma^2}n_B+\frac{3}{5}\frac{\hbar^2}{2m}\left(6\pi^2\right)^{2/3} n_F^{5/3} +\frac{g_{\rm dd} n_B^2}{2}+g_{\rm ad} n_B n_F+...$$

\noindent with $n_B=n_\downarrow$ is the number of bosonic dimers, $n_F=n_\uparrow-n_\downarrow$ is the number of free excess fermions and $g_{\rm dd}$  is the coupling constant describing s-wave dimer-dimer interactions and characterized by the scattering length $a_{\rm dd}=0.6 a$ \cite{petrov2004weakly,brodsky2006exact}. In contrast with the fermionic physics at play in the BCS region of the phase diagram where the gap maintains a perfect pairing, for $n_F\not = 0$, this equation of state describes a {\em polarized} superfluid. For weak enough repulsive interactions between bosons and fermions, the Bose-Fermi mixture is stable, while strong interaction close to the Feshbach resonance drives a first order phase separation~\cite{viverit2002ground,pilati2008psp,iskin2007,iskin2008} between a mixed Bose-Fermi phase and a fully polarized Fermi gas. The Fermi gas is made of majority fermions repelled from the Bose-Fermi mixture. A linear stability analysis yields the stability criterion

\be
n_F^{1/3} \le \frac{(6 \pi^2)^{2/3}}{12 \pi}  \frac{a_{\rm dd}}{a_{\rm ad}^2}
\label{Eqn3.3}
\ee

\noindent that is obtained from the positivity of the compressibility matrix $\partial_{\mu_i\mu_j}E$.
Using the values of $a_{\rm ad}$ and $a_{\rm dd}$ given above, we finally get that an atom/dimer phase separation  takes place for $1/k_{F\uparrow} a < 1.7$ when $n_\downarrow \to0$ (For comparison, the mean field prediction is $1/k_{F\uparrow} a \simeq 1.88$~\cite{parish2007finite}). This onset of a first order transition gives rise to a tricritical point where the order of the transition changes and that will be studied in more detail in the paragraph dedicated to finite temperature phenomena, see below.  The linear stability analysis also predicts a critical end-point at zero polarization
after which the bosonic superfluid can no longer sustain a finite
polarization. All additional majority fermions are therefore expelled.
The corresponding stability criterion is given by
\be
n_\uparrow = n_B \le \left( \frac{5}{4} \right)^2 \left(\frac{\hbar^2}{2 m} \right)^3 (6 \pi^2)^2 \frac{ g_{\rm dd}^2}{g_{\rm ad}^5},
\ee
for $n_\uparrow = n_\downarrow$. The critical end-point is thus located at $1/k_{F\uparrow} a \simeq 0.66$ in fairly good agreement with
the value  $1/k_{F\uparrow} a \simeq 0.53$ obtained with Fixed Node Monte-Carlo~\cite{pilati2008psp}.

The complete canonical phase diagram in the BEC-BCS crossover was first established from a Nozi\`eres-Shmitt-Rink analysis revealing its most important features~\cite{parish2007finite}. A more quantitative description was later provided by Fixed Node Monte-Carlo simulations \cite{pilati2008psp} and in Fig. \ref{Fig3.2}.b, we present a grand canonical  description of the phase diagram based on this work.

From dimensional analysis, the structure of the phase diagram can be encapsulated in two dimensionless numbers: $\eta=\mu_\downarrow/\mu_\uparrow$ which characterizes the spin imbalance and $\delta =\hbar/\sqrt{2m\mu_\uparrow}a$ which for an ideal gas would be equal to $1/k_Fa$ and thus measures the strength of interactions. Using Local density approximation, Fig. \ref{Fig3.2}.b can be used to predict the shape of the density profile in a trap. From equation $\mu_\uparrow(\bf r)=\mu_\uparrow^0-U(\bf r)$ and the definition of $\delta$, it is indeed possible to express $U(\bm r)$ as a function of $\delta (\bm r)$. Substituting in $\eta=\mu_\uparrow/\mu_\downarrow$, we then see that in a trap, the system follows a line $\eta=1-4\delta^2\bar\mu^0 m a^2/\hbar^2$, with $\bar\mu^{0}=(\mu_\uparrow^0-\mu_\downarrow^0)/2$. Interestingly, this equation does not depend on the actual shape of the trapping potential which only plays a role in the relationship between chemical potentials and atom numbers.

While the BCS side of the resonance has been thoroughly studied
experimentally, the predicted features of the phase diagram on the BEC
side have not yet been explored and await an experimental
confirmation. The increase in the atomic loss rate towards the BEC
region limits current experiments close to the unitary and BCS
regions.
In fact, even on the theoretical side, a clear understanding of the
bosonic/fermionic crossover at large polarization has not been
achieved yet. This is related to the question of whether  the
polaron/molecule transition is affected, or even preempted, by the
demixion mechanism described above, which leads to the coexistense of a
bosonic superfluid and a single-component
Fermi gas. Two scenarios can be envisioned. The first one is supported
by Quantum Monte-Carlo calculations. A first order transition occurs,
at~\cite{pilati2008psp} $1/k_{F\uparrow} a = 0.73$ in the impurity limit $n_{\downarrow}
\to 0$, where the polaronic normal phase separates into a polarized
paired superfluid immiscible with the remaining majority
fermions. This transition hides the polaron/molecule transition,
 that is located~\cite{prokof'ev08fpb} further at $1/k_{F\uparrow} a = 0.91(2)$.
The possible uncertainty in Monte-Carlo calculations and the proximity
of these two transition points allow to imagine an alternative
scenario where the two points indeed coincide. In that case, the
polaron/molecule transition would be accessible and immediately
followed by phase separation on the BEC side.
Interestingly, even the nature of the polaron/molecule phase
transition has not been unambiguously determined and a smooth
transition remains possible.

\subsection{Experimental characterization of the Fermi polaron.}

The review of imbalanced Fermi gases presented in the previous sections has shown that the behavior of the partially polarized normal phase could be essentially understood from the properties of the Fermi polaron. We show here that some of the experimental investigation tools developed for balanced Fermi gases can be applied to imbalanced situations. These studies confirm quantitatively some of the theoretical predictions presented above, in particular the spectrum of a single polaron or the equation of state of the partially polarized normal phase.

\subsubsection{Radiofrequency spectroscopy.}

The parameter $A$ that characterizes the transition between a fully polarized and a partially polarized normal phase was measured in Ref.~\cite{schirotzek2009ofp} using radio-frequency spectroscopy of a strongly imbalanced gas. When the imbalance is large enough (more specifically larger than $P_c\sim 0.75$ at unitarity), no superfluid core is observed, and all minority atoms are located in the normal component. The rf spectroscopy of spin down atoms therefore gives access to the energy shift $A E_{F\uparrow}$ induced by the presence of the majority Fermi sea. These results are presented in Fig. \ref{Fig3.2} and confirm remarkably well the theoretical approaches developed to describe the crossover. In addition, the measurement of the spectral weight of the radio-frequency resonances allowed for a characterization of the fermionic nature of the polaron and demonstrated that beyond $1/k_{F\uparrow} a\sim 0.74(4)$, the impurity was loosing its fermionic nature. This value is rather far from the molecule/polaron transition but is by contrast strikingly close to the prediction of the Fixed Node Monte-Carlo for the transition between the partially polarized superfluid and normal phases. It therefore suggests that, as discussed above, the single impurity molecule/polaron transition is preempted by the phase separation between the ideal Fermi gas and the polarized molecular superfluid.

\subsubsection{Dynamics and effective mass.}

To measure the effective mass of the polaron, it is necessary to study the dynamics of an imbalanced Fermi gas. As proposed in \cite{lobo2006nsp}, this can be done by the study of the collective modes of the mixture. Indeed, the energy of a trapped polaron is given by the equation

\be
E(\bm r,\bm p)=A E_{F\uparrow}(\bm r)+\frac{p^2}{2m^*}+U(\bm r)=A E_{F\uparrow}(\bm 0)+\frac{p^2}{2m^*}+(1-A)U(\bm r),
\ee

\noindent where we have used the LDA condition $E_{F\uparrow}(\bm r)=E_{F\uparrow}(0)-U(\bm r)$. In the case of a harmonic trapping $U(\bm r)=\sum_{i=x,y,z} m\omega_i^2 x_i^2/2$, we see that the  trapping potential of the polaron is still harmonic, with an effective frequency

\be
\omega_i^*/\omega_i=\sqrt{(1-A)m/m^*}.
\label{Eqn3.1}
\ee

Using the magnetic field dependence of the scattering length, it is possible to excite selectively the minority atoms by canceling the interatomic interactions. Such an experiment was performed in Ref.~\cite{nascimbene2009pol}. The study of the axial breathing mode of an elongated imbalanced Fermi gas showed that at unitarity the oscillation frequency was about 1.17(2) times bigger than that of an ideal gas. From this measurement, Eq. (\ref{Eqn3.1}) and the value $A=-0.6$ obtained from theory or rf measurement yield $m^*=1.17(10) m$, in agreement with theories of the polaron presented previously.

\subsubsection{Collective behavior.}

\begin{figure}
\centerline{\includegraphics[width=0.5\columnwidth]{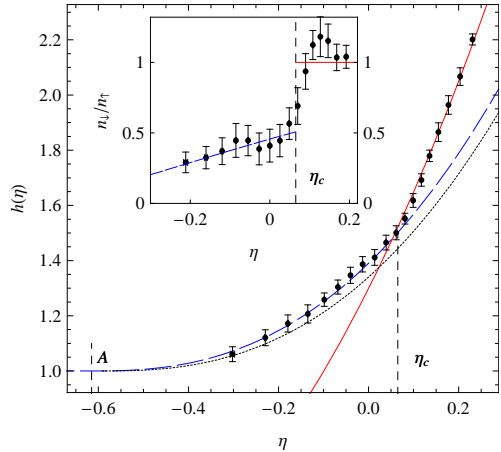}}
\caption{Equation of state of a unitary imbalanced Fermi gas in the grand canonical ensemble. The mixed normal component spans $\eta=[A,\eta_c]\sim [-0.6,0.065]$. The dashed line corresponds to the polaron Fermi liquid Eq. \ref{Eqn3.2}, while the dotted line is the prediction of the Monte-Carlo prediction \cite{lobo2006nsp}. Above $\eta=\eta_c$, the equation of state is well described by that of unitary gas with $\xi_s=0.42(1)$. Inset: density ratio $n_\downarrow/n_\uparrow$ calculated from the thermodynamic identity $n_i=\partial_{\mu_i} P$. The critical value $\eta=\eta_c$ separates a superfluid with $n_\uparrow=n_\downarrow$ and a partially polarized normal phase. Figure from \cite{nascimbene2009eos}.}
\label{Fig3.3}
\end{figure}

In the polaron picture, the partially polarized normal component is
described by the equation of state of a  mixture of two ideal Fermi
gases (Eq. \ref{Fig3.2}). Using the general method described in
section \ref{SectionDensityProfile}, it is possible to directly
measure this grand-canonical equation of state by the analysis of
absorption imaging pictures, as displayed in Fig. \ref{Fig3.3}.a
\cite{nascimbene2009eos}. A remarkable agreement is thus obtained
between experiments and the Fermi liquid model elaborated  for the properties of the Fermi-polaron quasi-particles.

This procedure can be extended to the full crossover and, except in the far BEC region of the phase diagram where the polaron picture breaks down the equation of state (\ref{Eqn3.2}) agrees  with Monte-Carlo simulations \cite{bertaina2009dpp} as well as experimental data \cite{navon2010Ground}. This observation confirms that Fermi-polarons are indeed fermions whose effective parameters $A$ and $m^*$  are well understood using the variational model described above.

\subsubsection{The bosonic sector}
The experimental exploration of bosonic region of the crossover is presented in \cite{shin2008rsi}. This work first confirms the disappearance of the polaron phase for $1/k_{F1}>0.75$, in agreement with Monte-Carlo simulations. The analysis of the cloud density profile furthermore demonstrates that beyond this threshold the system behaves like the Bose-Fermi mixtures described above. Quantitative comparison between theory and experiment allowed for the measurement of the dimer/dimer and atom/dimer scattering lengths confirming the expected values $a_{dd}=0.6 a$ and $a_{ad}=1.2a$. Interestingly, experimental data showed small deviations from the bosonic mean-field behavior that could be explained by the introduction of beyond-mean-field Lee-Huang-Yang term \cite{lee1957eigenvalues}.

\section{Finite Temperature phase-diagram}

\begin{figure}
   \begin{minipage}[c]{0.46\linewidth}
      \includegraphics[width=\linewidth]{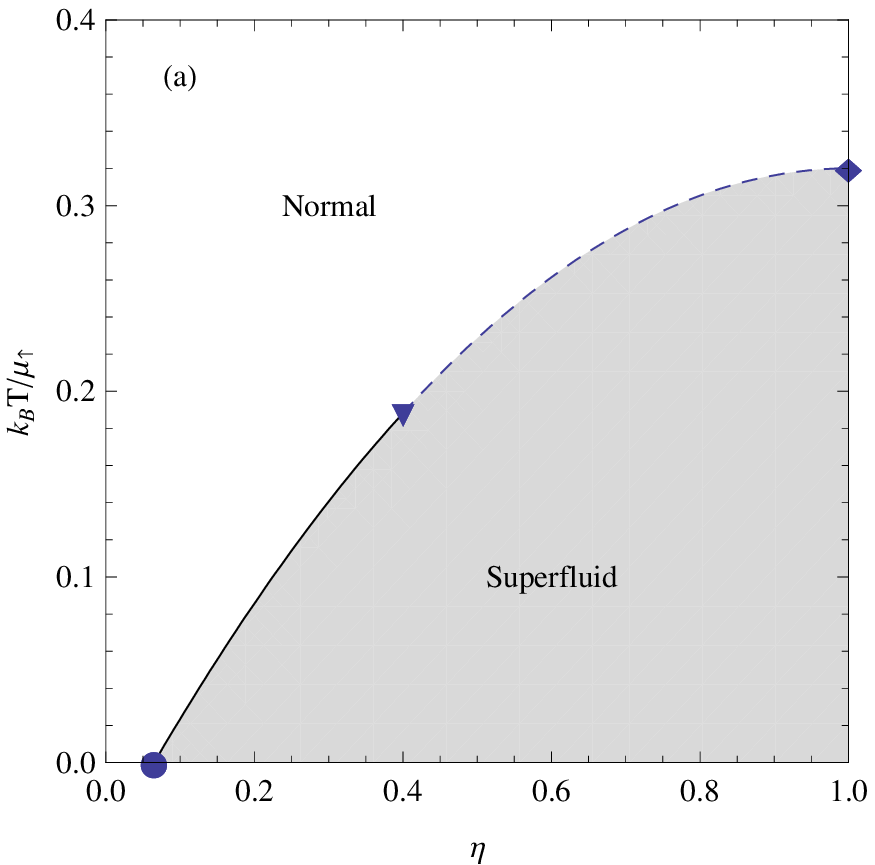}
   \end{minipage} \hfill
   \begin{minipage}[c]{.46\linewidth}
      \includegraphics[width=\linewidth]{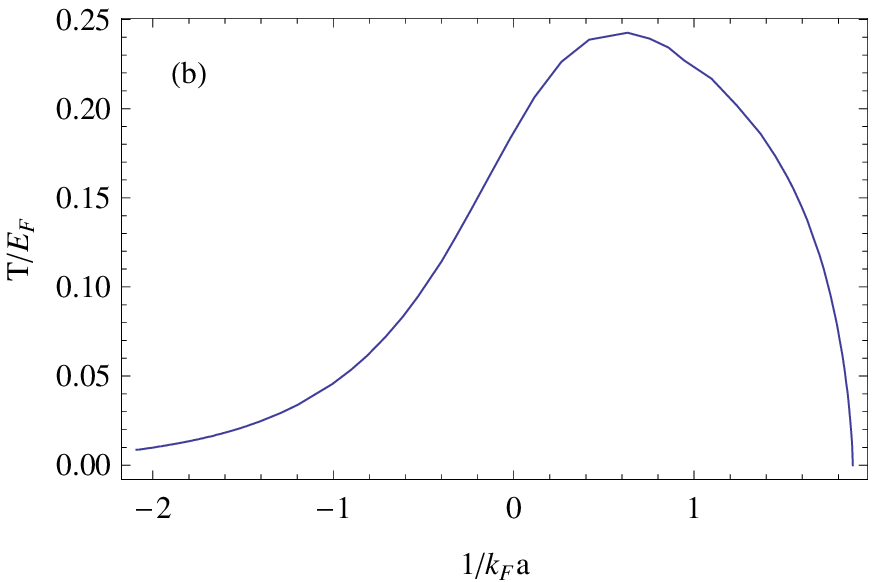}
   \end{minipage}
\caption{(a): Sketch of the finite temperature phase diagram of the unitary
  imbalanced Fermi gas. The phase transition being first order at zero
  temperature (circle) and second order at zero polarization (diamond)
  implies the existence of a tricritical point located at $(\eta_{\rm
    t},T_{\rm t})$, (triangle). Above this point, the critical line is
  second order (Dashed line), and first order below (Solid line). (b):
  Tricritical temperature in the BEC-BCS crossover. In the far BCS
  limit, $T_t$ is proportional to $T_c$ and is thus exponentially
  small. At $1/k_{F\uparrow} a=1.7$, the tricritical temperature
  vanishes and coincides with the zero-temperature demixing threshold
  from Eq. \ref{Eqn3.3} (diamond) (Data from \cite{parish2007finite}, courtesy of M. Parish).}
\label{Fig4.1}
\end{figure}

The  phase diagram discussed in the previous section displays intriguing features when extended to the finite temperature regime. Indeed, thermal fluctuations reduce the value of the gap $\Delta$ of the excitation spectrum and, according to the CC argument, weaken the robustness of the superfluid against spin polarization. We thus expect the  normal-superfluid transition to follow a critical line $\eta_c(T_c)$ in the $(\eta, k_B T/\mu_\uparrow)$ plane, starting from $\eta_c(0)=0.065$ as discussed earlier, and ending at $\eta (T_c)=1$, for some critical temperature $T_c$ for which the superfluid displays no longer any resistance to polarization. Since the phase transition takes place at zero imbalance, we identify $T_c$ with the usual superfluid/normal second order transition that was was recently located at $k_B T_c/\mu=0.32$ at unitarity \cite{burovski2006critical,zwierlein2006dos,nascimbene2009eos}. In the BCS weakly interacting regime, the position of this transition can be obtained from BCS theory. Remarkably, while at zero temperature the transition is first order, as discussed above, we expect it to become second order at zero imbalance. This suggests the existence of a tricritical point somewhere in between as seen in Fig. (\ref{Fig4.1}) which outlines the main  features of the finite temperature phase diagram of a unitary Fermi gas.

Interestingly, the Bose/Fermi  behavior of the polarized Fermi gases on the BEC side of the resonance has similarities
with $^3$He/$^4$He mixtures~\cite{GrafPhase1967} and the phase diagrams thus share common features.
In particular, a first order phase transition between phases with different isotopic ratios were observed at low temperature in $^3$He/$^4$He. Since this phenomenon forms the basis of dilution refrigerators, the physics  of helium mixtures is of the highest practical interest and their thermodynamic properties around tricritical points have been studied extensively using Renormalization Group approaches in the early 70's \cite{RidelTricritical1972,RiedelScaling1972,WegnerLogarithmic1973}. In this context, it was demonstrated that, within logarithmical corrections, the critical exponents in the vicinity of the second order/first order transition in 3D were given exactly by Landau theory.




In the case of ultra-cold gases, renormalization group calculations have  been used to derive fairly accurately the position of the tricritical point at unitarity  \cite{gubbels2008ren}, and yielded $T_{\rm t}/T_{F\uparrow}=0.09$. This value is in close agreement with experimental observations that measured its position by  the disappearance of the density jump at unitarity associated with the first order nature of the normal-superfluid transition a large imbalance \cite{shin2008pd}. A semi-quantitative theoretical extension  to the BEC-BCS crossover has been proposed in \cite{parish2007finite}. Using Nozi\`eres-Shmitt-Rink approximation, it is possible to show that the temperature of the tricritical point follows a curve which connects the unitary tricritical point to the zero-temperature demixing threshold (Eq. \ref{Eqn3.3}) where, as mentioned earlier, the normal/superfluid transition goes also from first to second order.

\begin{figure}
\centerline{\includegraphics[width=\columnwidth]{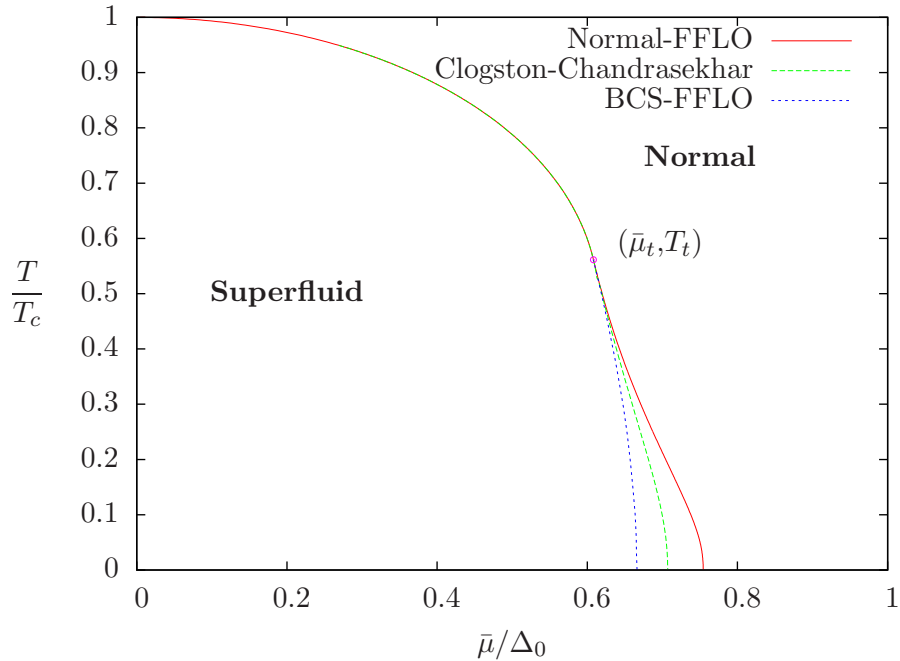}}
\caption{Mean field phase diagram including FFLO phases (BCS
  limit). This graph generalizes the phase diagram of
  Fig. \ref{Fig4.1} to the case where the FFLO phase is taken into account. The tricritical point corresponds also to the disappearance of the inhomogeneous superfluid.}
\label{Fig4.2}
\end{figure}

The tricritical point bears an intriguing connexion with FFLO phases
(Fig. \ref{Fig4.2}). Indeed, in the weak coupling BCS regime, the domain with
FFLO phases shrinks as the temperature is increased from zero:
the first order FFLO-normal state transition line and the second order homogeneous(BCS)-FFLO
transition line meet exactly at the tricritical point~\cite{sarma1963,combescot2002},
located at $T_t/T_c \simeq 0.561$, $\bar{\mu}_t/\Delta_0 \simeq 0.608$
for weak coupling, where $T_c$ is the critical temperature in the balanced case.
Above the tricritical point, the superfluid-normal transition is the standard second order transition
to  the homogeneous BCS phase. Interestingly, the Clogston-Chandrasekhar limit is always preempted by
the FFLO transition and also ends at the tricritical point. This feature is specific to the weak coupling limit~\cite{burkhardt1994}
and other scenarios can be envisioned at stronger coupling~\cite{gubbels2009}.

\section{Conclusion}
Experimental results obtained in the past years at MIT, Rice, and  ENS have offered a  simple picture of the phase diagram of an imbalanced Fermi gas with attractive interactions. Quite surprisingly, the main features can be captured with a remarkable precision by models based on simple concepts such as Fermi Liquid theory or polaron physics.  Despite these undeniable successes, many open questions remain and will be explored in the near future. The most serious one is the discrepancy observed between Rice results and those of MIT and ENS. The absence of a normal mixture in Rice's observations is still an unsettled issue and several explanations have been proposed, among which the role of trap anisotropy, atom number \cite{desilva2006stu,ku2009finite} or evaporation \cite{ParishEvaporative2009}. However, ENS experiments were performed in a regime close to that of Rice, and do shows features very similar to those of MIT.

 The second point is related to the existence of a FFLO phase, and more generally that of a the existence of a partially polarized superfluid. Although no systematic experimental study of this question has been performed in 3D, the remarkably good agreement between theory and experiment suggests that the FFLO phase can occupy only a small region of the phase diagram and/or do not influence much the macroscopic properties of the cloud. It was nevertheless  suggested that, in the spirit of the celebrated Kohn-Luttinger theorem \cite{Kohn1965New}, at low temperature the polaron gas could form a p-wave superfluid \cite{bulgac2006ipw}, but up to now, no experimental signature of this new phase was reported.  It was known from earlier works~\cite{machida1984,buzdin1983} that a reduced dimensionality, in particular in 1D, improves the Fermi surface nesting and therefore favors the emergence of FFLO-like features. In the last few years, a considerable theoretical effort was devoted to understand and predict the existence and fingerprints of the FFLO polarized superfluid state in 1D, combining bosonization approaches~\cite{yang2001,Zhao2008}, Bethe-Ansatz calculations of the Yang-Gaudin model~\cite{Orso2007Attractive,Hu2007Phase,Kakashvili2009}, Monte-Carlo~\cite{Batrouni2008,Casula2008,roscilde2009quantum} and DMRG studies~\cite{feiguin2007,Tezuka2008,Machida2008,rizzi2008,luescher2008,Heidrich-Meisner2010,masaki2010} of the attractive Hubbard model or of the Yang-Gaudin model, and a spin-density-functional theory~\cite{gao2008}.  Recently, very promising results where obtained in this direction by the group of Rice who trapped ultra-cold fermions in elongated quasi-1D tubes \cite{liao2009spin}. The study of the density profiles was in good agreement with theoretical predictions displaying FFLO-like behavior.


Third, despite important progress in the understanding of static properties of imbalanced Fermi gases, very little is known on their dynamic behavior.  One challenge is the  understanding of the crossover between the hydrodynamic and collisionless regimes at low and high polarizations respectively. First experimental evidences were presented in \cite{nascimbene2009pol}, but a unifying theoretical description is still missing, despite some attempts using Fermi-Liquid theory \cite{bruun2008collisional,stringari2009density}, or the resolution of coupled Boltzmann/Euler equations \cite{lazarides2008collective}.

Finally an interesting perspective is offered by the recent realization of quantum degenerate Lithium-Potassium Fermi mixtures \cite{wille2008exploring,taglieber2008quantum}.  In particular, recent results on the width of the Feshbach resonances \cite{ticke2010broad} and the stability of the $^6$Li/$^{40}$K mixture \cite{spiegelhalder2009collisional}  offer the opportunity to explore experimentally a new generalization of the BCS superfluidity mechanism where mismatch between Fermi surfaces is created by mass imbalance, leading to a phase diagram even richer than the one presented here~\cite{gubbels2009,bausmerth2009,gezerlis2009,parish2007pfc,mathy2010polarons,Baranov2008Superfluid}.

\ack We are grateful to C. Salomon, S. Nascimb\`ene and N. Navon for fruitful discussions. FC acknowledges support from  EU (ERC project Ferlodim), IFRAF, ANR (Project FABIOLA) and Institut Universitaire de France.
\section*{References}
\bibliographystyle{unsrt}
\bibliography{bibliographie}
\end{document}